\documentclass{blois}

\newcommand{\lsim}
{\;\raisebox{-.3em}{$\stackrel{\displaystyle <}{\sim}$}\;}

\newcommand{\GeV}{\unskip\,\mathrm{GeV}}

\newcommand{\TeV}{\unskip\,\mathrm{TeV}}

\def\mathswitch#1{\relax\ifmmode#1\else$#1$\fi}
\def\mathswitchr#1{\relax\ifmmode{\mathrm{#1}}\else$\mathrm{#1}$\fi}
\def\mathswitchit#1{\relax\ifmmode{#1}\else$#1$\fi}

\newcommand{\PW}{\mathswitchr W}
\newcommand{\PZ}{\mathswitchr Z}

\newcommand{\Pg}{\mathswitchr g}
\newcommand{\PH}{\mathswitchr H}

\newcommand{\Pt}{\mathswitchr t}
\newcommand{\Pp}{\mathswitchr p}

\newcommand{\MW}{\mathswitch {M_\PW}}

\newcommand{\MH}{\mathswitch {M_\PH}}

\newcommand{\Mt}{\mathswitch {m_\Pt}}

\def\be{\begin{equation}}
\def\ee{\end{equation}}
\def\bea{\begin{eqnarray}}
\def\eea{\end{eqnarray}}

\newcommand{\Photo}{\vspace*{.5em}\includegraphics[height=35mm,angle=90]{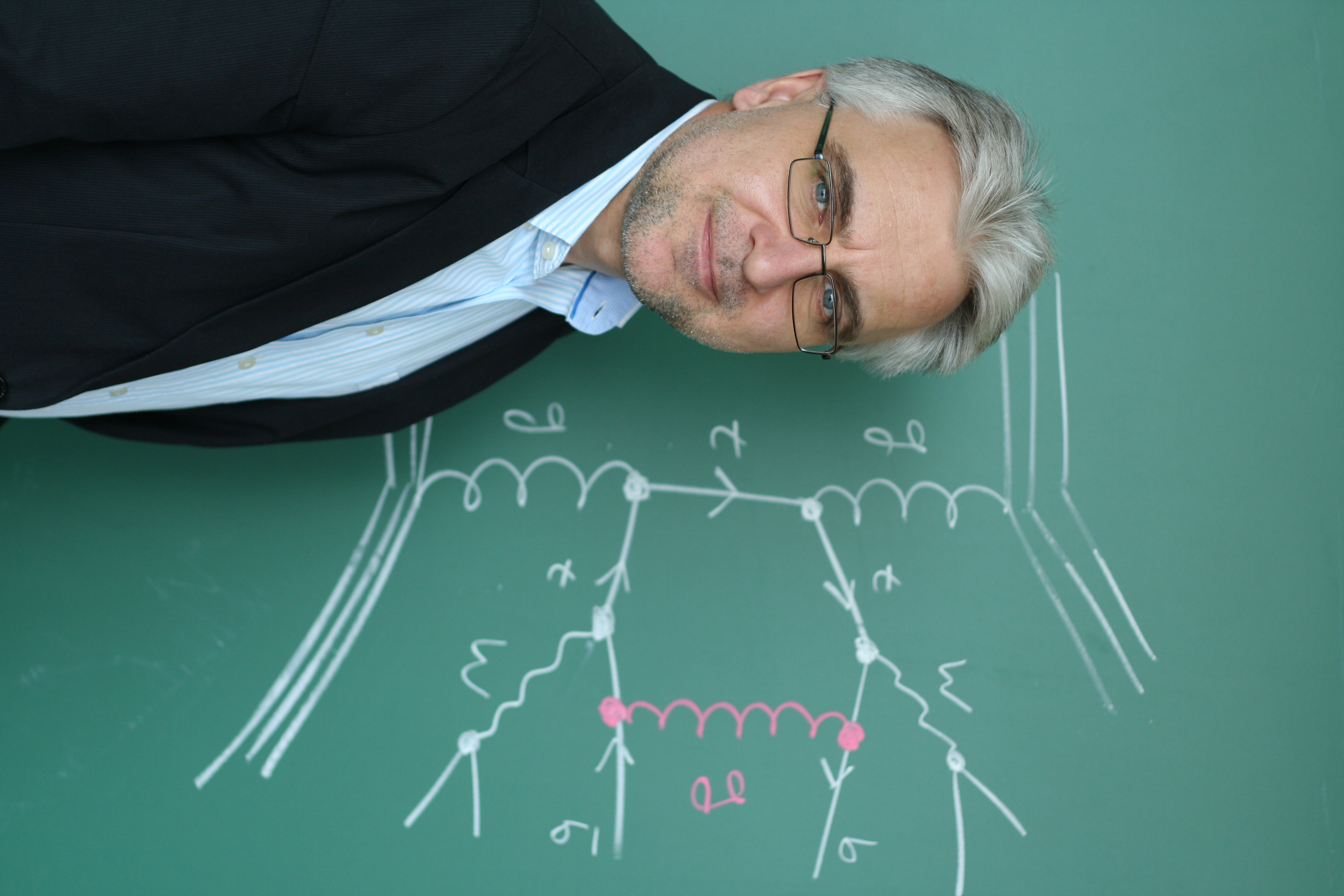}}

\begin{document}
\vspace*{4cm}
\title{STANDARD MODEL THEORY}

\author{STEFAN DITTMAIER}

\address{Albert-Ludwigs-Universit\"at Freiburg, Physikalisches Institut, \\
79104 Freiburg, Germany}

\maketitle\abstracts{
Recent progress in the field of precision calculations for Standard Model processes
at the LHC is reviewed, highlighting examples of weak gauge-boson and Higgs-boson
production, as discussed at the 27th Rencontres de Blois, 2015.
}

\section{Introduction}

After Run~1 of the LHC, the Standard Model (SM) is in better shape than
ever in describing practically all phenomena in high-energy particle physics.
The search for new physics, thus, has to proceed with precision at the highest possible
level, in order to reveal any possible deviation from SM predictions.
To this end, both QCD and electroweak (EW) corrections have to be included in 
cross-section predictions.

The field of perturbative precision calculations has experienced tremendous
progress in recent years in various directions. 
Next-to-leading-order (NLO) QCD calculations have been automated up to
particle multiplicities of roughly 4--6 (depending on the complexity of the
process) upon combining multi-purpose Monte Carlo generators or integrators with 
automated tools for the generation of multi-leg one-loop matrix elements.
For NLO EW corrections the automation is in progress as well.
At the next-to-next-to-leading-order (NNLO) level more and more
complete QCD calculations have been completed for various $2\to2$ particle
scattering processes, and for $2\to1$ processes even first results at
next-to-next-to-next-to-leading order (NNNLO) are presented.
This progress in fixed-order calculations goes in parallel with new achievements
in the resummation of leading corrections to all perturbative orders.
On the one hand, analytic QCD resummations were continuously pushed to 
higher and higher levels. 
On the other hand, fixed-order calculations were matched to QCD parton showers,
where the NLO level meanwhile follows standard procedures and the frontier
moved on to NNLO~\cite{Hoeche:2014aia}.
In cases where the matching of NLO corrections with parton showers does not yet
deliver sufficient precision in the first few jet multiplicities,
improvements are often obtained upon merging NLO calculations for
producing a specific event topology together with $n=0,1,2,...$ jets,
thereby carefully avoiding double-counting of jet activity.

In this short review, some highlights in the recent progress in fixed-order
calculations are summarized. In detail, advances in precision calculations
for the production of W/Z~bosons~+~jets and of weak gauge-boson pairs are discussed
as well as recent results on the Higgs-boson production rate and the corresponding
transverse-momentum spectrum.

\section{Weak-gauge-boson production}

\subsection{W/Z production in association with hard jets}

The production of W or Z bosons in association with hard jets represents an
important class of standard processes at the LHC, both as testground for
jet dynamics in QCD and as background process to many searches.
The corresponding SM predictions experienced an enormous boost in precision
in recent years. 

\begin{figure}
\centerline{\includegraphics[width=0.6\linewidth]{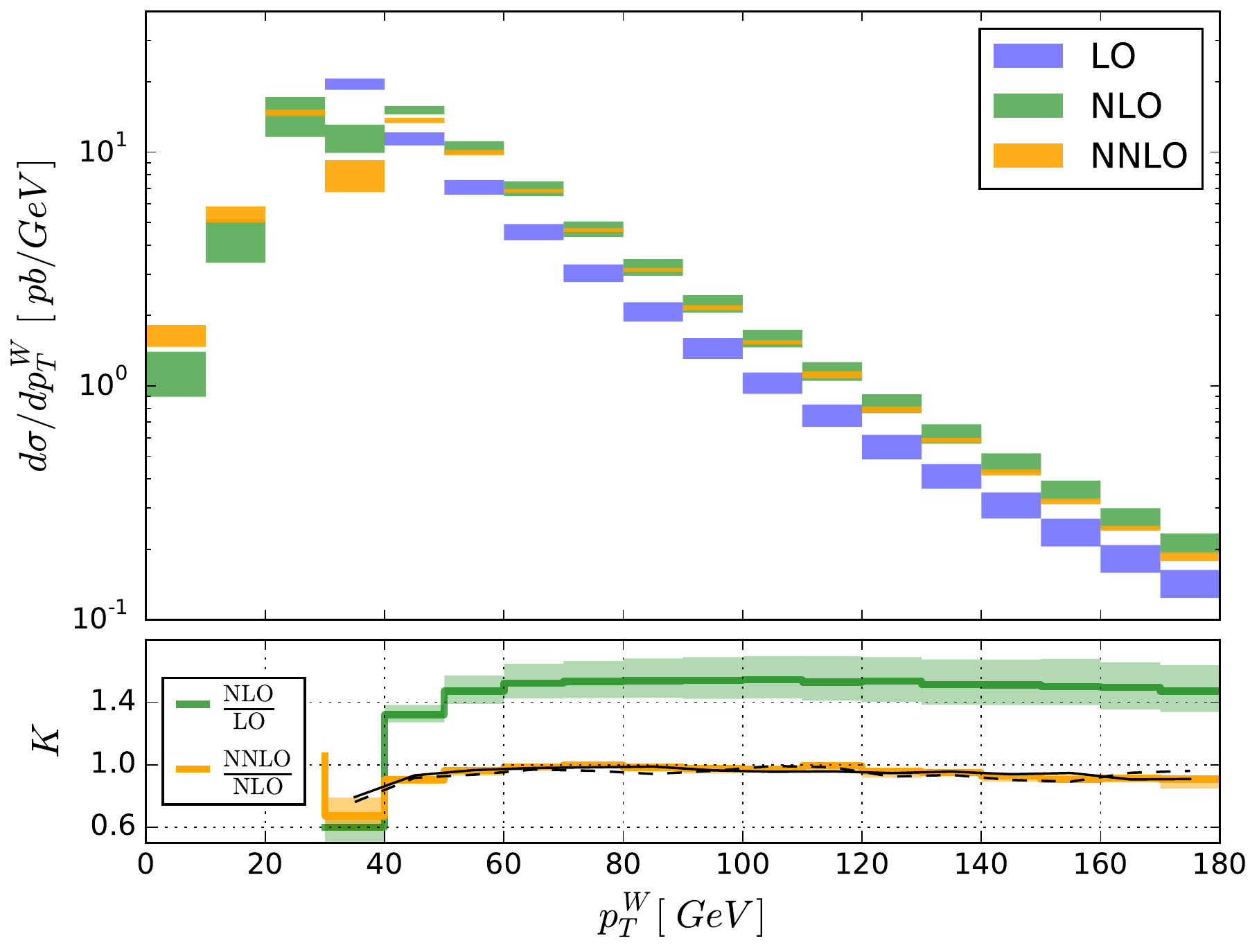}}
\caption[]{NNLO QCD prediction for the transverse-momentum ($p_{\mathrm{T}}$)
spectrum of the W~boson in
$\PW^+(\to\ell\nu)+$jet production at the LHC at $8\TeV$
and corresponding $K$~factors 
(taken from Boughezal et al.~\protect{\cite{Boughezal:2015dva}}).
The factorization and renormalization scales are set to 
$\mu=\MW$ and simultaneously varied by a factor of~2 up and down in the shown
uncertainty bands.
}
\label{fig:W+1j}
\end{figure}

After an effort over many years, very recently the QCD-based predictions 
for W/Z+1jet were pushed to the NNLO level~\cite{Boughezal:2015dva,Ridder:2015dxa}.
As one of the central results, 
Fig.~\ref{fig:W+1j} shows the transverse-momentum spectrum of the W~boson in
$\PW^+(\to\ell\nu)+$jet production at the LHC at $8\TeV$ at LO, NLO, and 
NNLO~\cite{Boughezal:2015dva}.
While fixed-order predictions generally are not able to describe the range of
low $p_{\mathrm{T}}^\PW$, for intermediate and large $p_{\mathrm{T}}^\PW$
the perturbative series shows nice convergence.
For $p_{\mathrm{T}}^\PW$ in the range $50\GeV\lsim p_{\mathrm{T}}^\PW \lsim 180\GeV$,
the QCD corrections are $\sim40\%$ at NLO, but only a few percent at NNLO,
reducing the residual scale uncertainty from $\sim20\%$ at NLO to only
few percent at NNLO.
The pattern of the NNLO QCD corrections to Z+1jet production is very 
similar~\cite{Ridder:2015dxa}.
To bring the overall theory uncertainty to the level of the missing QCD corrections
of some percent, in particular, the EW corrections at the NLO level
have to be taken into account. They are known not only for on-shell
W/Z bosons~\cite{Kuhn:2005az},
but also including all off-shell and decay effects~\cite{Denner:2009gj}.

In the completion of the NNLO QCD calculations the major obstacle
was the extraction of infrared (IR) singularities from the real double-emission
and the real--virtual contributions and their cancellation against their
virtual counterparts.
The major breakthrough in the calculation
for W+1jet production~\cite{Boughezal:2015dva}
was the invention of a new technique called
``jettiness subtraction''~\cite{Boughezal:2015eha},
where the soft/collinear IR singularities
are isolated by some cut on the ``jettiness''~\cite{Stewart:2010tn} ${\cal T}_N$ of the events,
\begin{equation}
{\cal T}_N \;=\;
\sum_k \min_i\left\{\frac{2p_i\cdot q_k}{Q_i}\right\}.
\end{equation}
The procedure works as follows:
First the number $N$ of jets is determined with any jet algorithm,
thereby defining $N$ light-like reference momenta $p_i\;$ 
(+2 beam momenta for $\Pp\Pp$ collisions).
Then ${\cal T}_N$ is calculated from the sum over all parton momenta $q_k$,
taking appropriate scales $Q_i$ to characterize the hardness of the jets.
The limit ${\cal T}_N\to0$ corresponds to exactly $N$ resolved jets and is
independent of the jet algorithm.
The phase space can, thus, be partitioned into regions with
${\cal T}_N<{\cal T}_N^{\mathrm{cut}}$ and
${\cal T}_N>{\cal T}_N^{\mathrm{cut}}$, where the former isolates the IR-singular
regions where factorization properties of the amplitudes can be used to 
analytically integrate over the singular degrees of freedom.
Since the structure of this procedure is rather generic, the technique,
which has been suggested in different 
variants~\cite{Boughezal:2015eha,Gaunt:2015pea},
should find more applications in forthcoming NNLO calculations.
We note in passing that
the NNLO calculation for Z+1jet production~\cite{Ridder:2015dxa}
was based on a different well-established
approach for the treatment of IR singularities known as
``antenna subtraction''~\cite{GehrmannDeRidder:2005cm}.

\begin{figure}
\includegraphics[bb= 0 187 270 516,clip,width=0.3\textwidth]{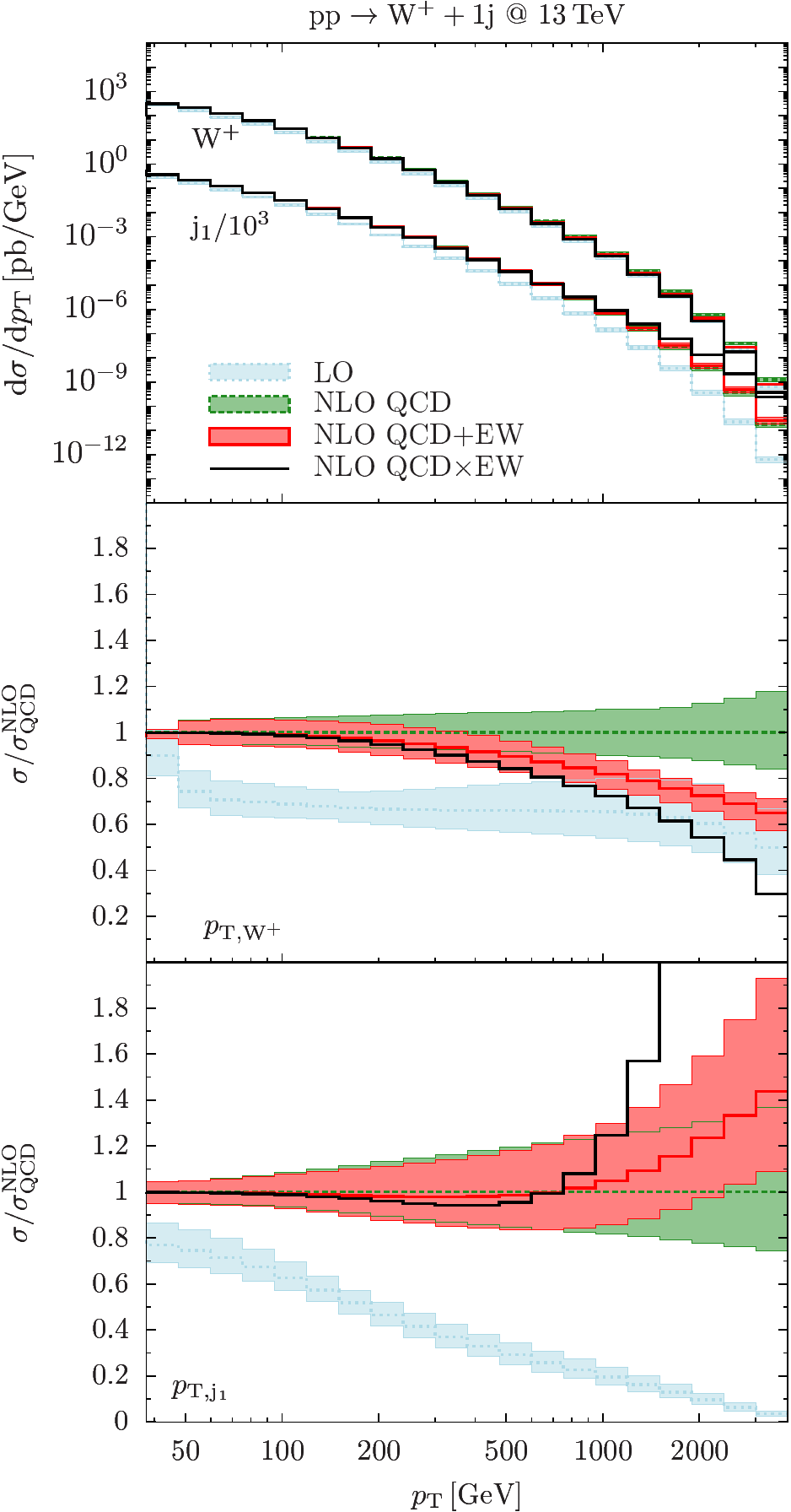} 
\hspace{1em}
\includegraphics[bb= 0 345 270 674,clip,width=0.3\textwidth]{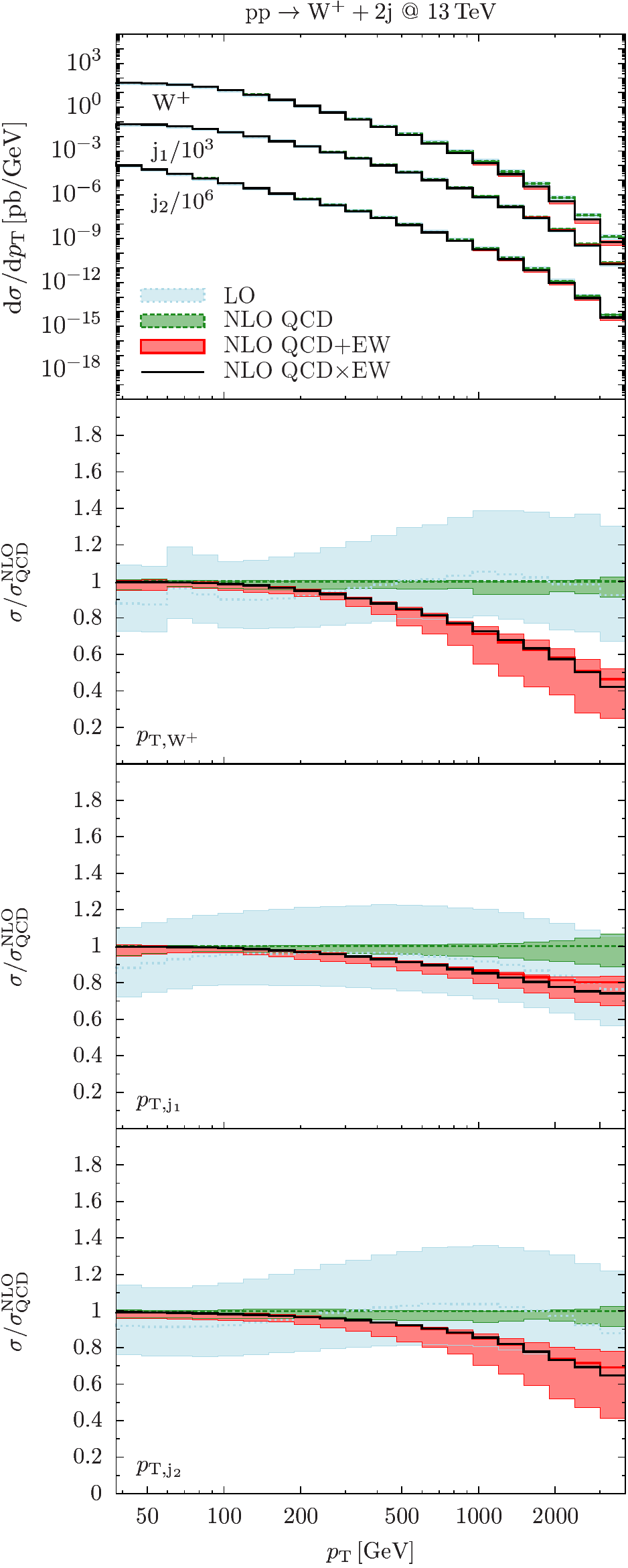} 
\hspace{1em}
\includegraphics[bb= 0 501 270 830,clip,width=0.3\textwidth]{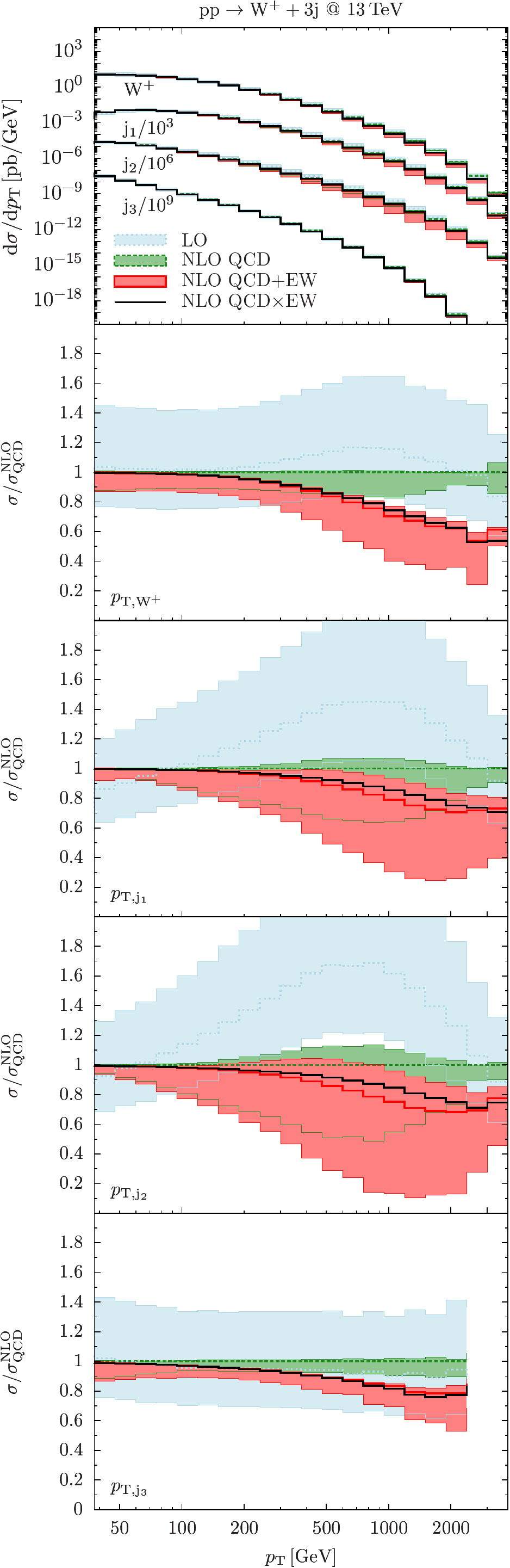} 
\\
\includegraphics[bb= 0 0 270 28 ,clip,width=0.3\textwidth]{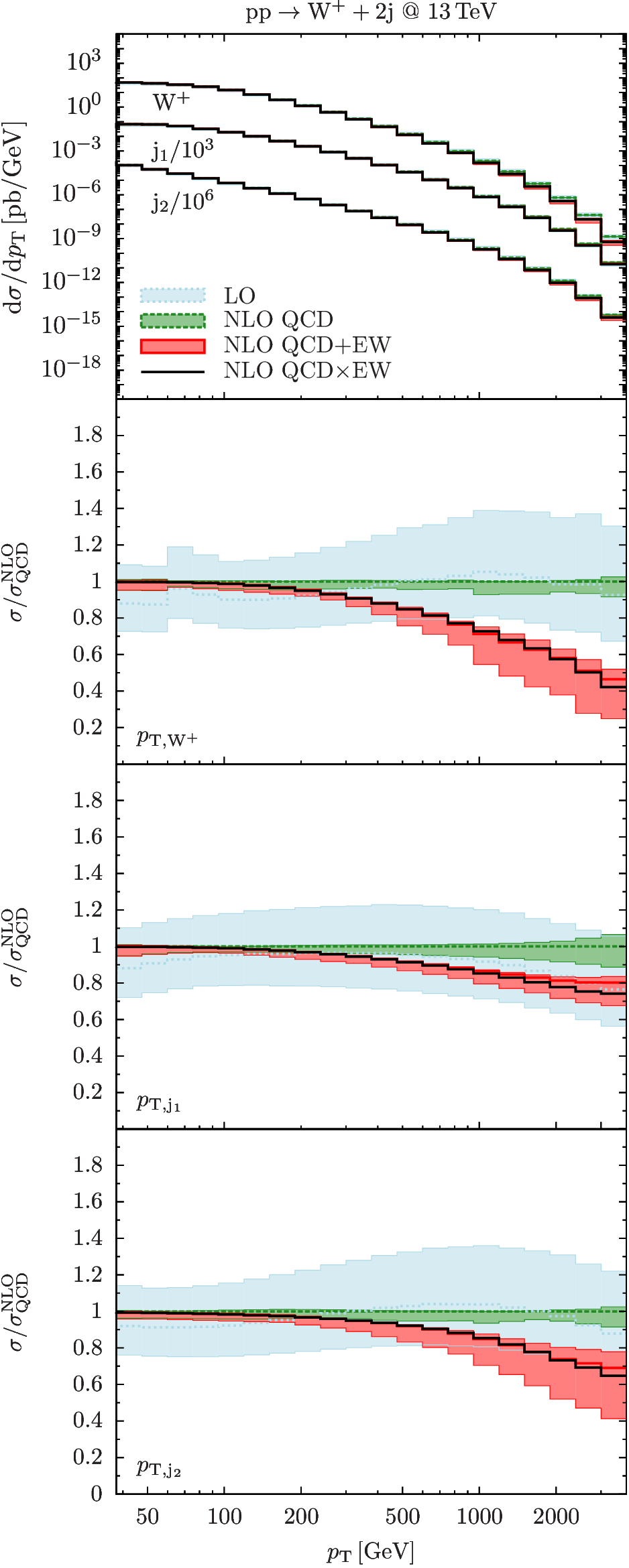}
\hspace{1em}
\includegraphics[bb= 0 0 270 28,clip,width=0.3\textwidth]{ppwxjj_incl_pTall_gp}
\hspace{1em}
\includegraphics[bb= 0 0 270 28,clip,width=0.3\textwidth]{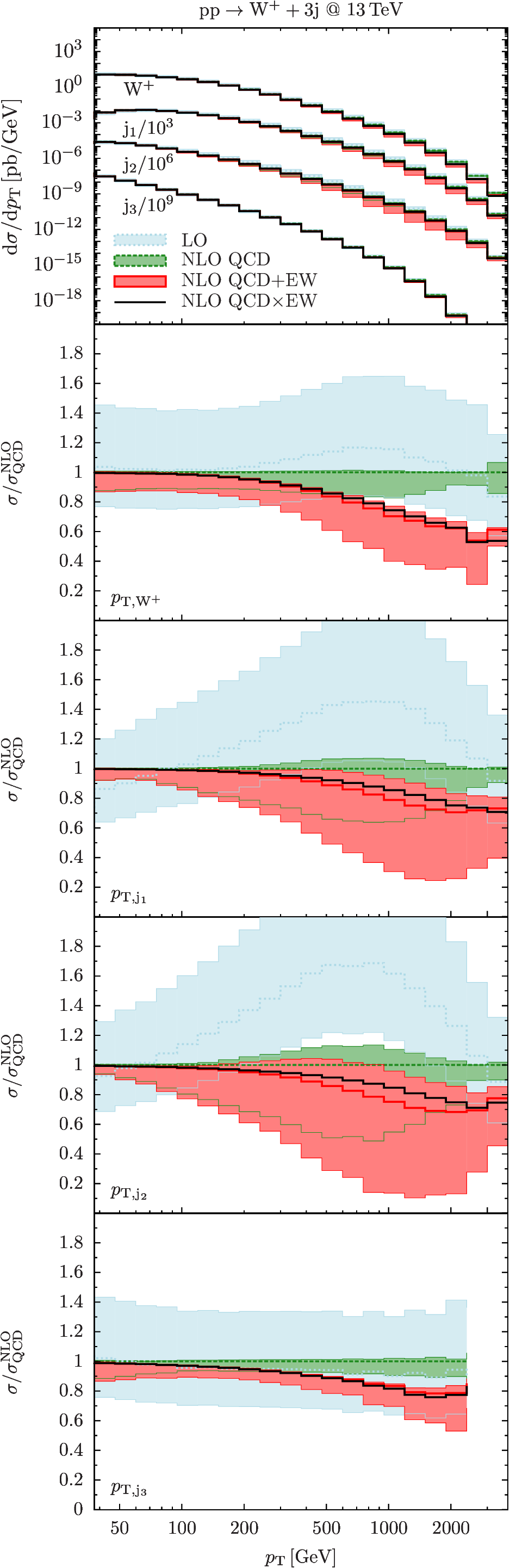}
\caption[]{Transverse-momentum distributions for the production of
$\PW^++1,2,3$jets in LO and including NLO QCD and EW corrections
 (taken from Kallweit et al.~\cite{Kallweit:2014xda}).
}
\label{fig:W+njets}
\end{figure}

Turning to higher jet multiplicities, the NLO level represents the state of the art. 
NLO QCD predictions even went up to 
3~\cite{Ellis:2009zw},
4~\cite{Berger:2010zx},
and 5~\cite{Bern:2013gka}
jets in association with a W or Z boson, at least in leading-colour approximation.
Concerning EW corrections, NLO predictions were recently calculated for
Z+2jet production~\cite{Denner:2014ina}
(with Z-boson decays and off-shell effects)
and W+1,2,3jet production~\cite{Kallweit:2014xda} 
(for on-shell W~bosons).
Exemplarily Fig.~\ref{fig:W+njets} compares the NLO QCD and EW corrections to the W-boson
transverse-momentum distribution for the three different jet multiplicities.
Normalizing the relative corrections to NLO QCD, the plots reflect the well-known
``giant QCD K~factor'' for W+1jet, but QCD corrections at the $10\%$ level for
2~and 3~jets. The scale uncertainty is roughly $\sim10\%$ ($\sim20\%$) for
$p_{\mathrm{T},\PW}$ up to $500\GeV$ ($2\TeV$).
The EW corrections show the generic feature to grow large and negative for
increasing $p_{\mathrm{T},\PW}$, reaching tens of percent in the TeV range,
as consequence of the EW Sudakov logarithms at high scales, which are
due to the soft/collinear exchange of virtual W/Z~bosons.
There are, however, other mechanisms as well that potentially cause
sizeable EW corrections at higher energies, such as interferences with diagrams
where jets are initiated by EW particles or channels with photons in the
initial state. Details about those effects, including also other observables,
are discussed by Kallweit et al.~\cite{Kallweit:2014xda}.
Note also the difference between the two variants of combining QCD and EW
corrections based on mere addition (red curves) or assuming factorization
(black curves). For W+1jet production the difference becomes large, since both
QCD and EW corrections are large. While there are good arguments 
favouring factorization, since large universal QCD and EW effects factorize,
ultimately the uncertainty can only be resolved by an NNLO QCD--EW calculation.

For single-W/Z production a strategy for calculating NNLO QCD--EW corrections
was worked out in Ref.~\cite{Dittmaier:2014qza}
for W/Z~bosons near their mass shell.
Already this relatively simple example at NNLO reveals~\cite{Dittmaier:2014koa}
that the
success of a naive factorization of relative NLO QCD and NLO EW correction
factors is rather limited, since the naive factors do not account
for the accumulation of jet and photon recoils in the double-real corrections.

\subsection{Electroweak gauge-boson pair production}

We turn to pair production processes of EW gauge bosons
where NNLO QCD predictions experienced a breakthrough in the previous two years
with completed calculations for 
$\PZ/\PW\gamma$~\cite{Grazzini:2013bna,Grazzini:2015nwa},
$\PZ\PZ$~\cite{Cascioli:2014yka},
$\PW\PW$~\cite{Gehrmann:2014fva} based on on-shell W/Z gauge bosons,
and recently for $\PZ\PZ$~\cite{Grazzini:2015hta}
even for off-shell Z~bosons.
The issue of extracting and cancelling IR singularities in these
calculations was accomplished by so-called
``$q_{\mathrm{T}}$~subtraction''~\cite{Catani:2007vq},
which exploits the fact that QCD emission in processes with colour-neutral
final states can be isolated in the limit where jets have small
transverse momentum.

\begin{figure}
{\includegraphics[bb=18 190 594 590,width=0.5\textwidth]{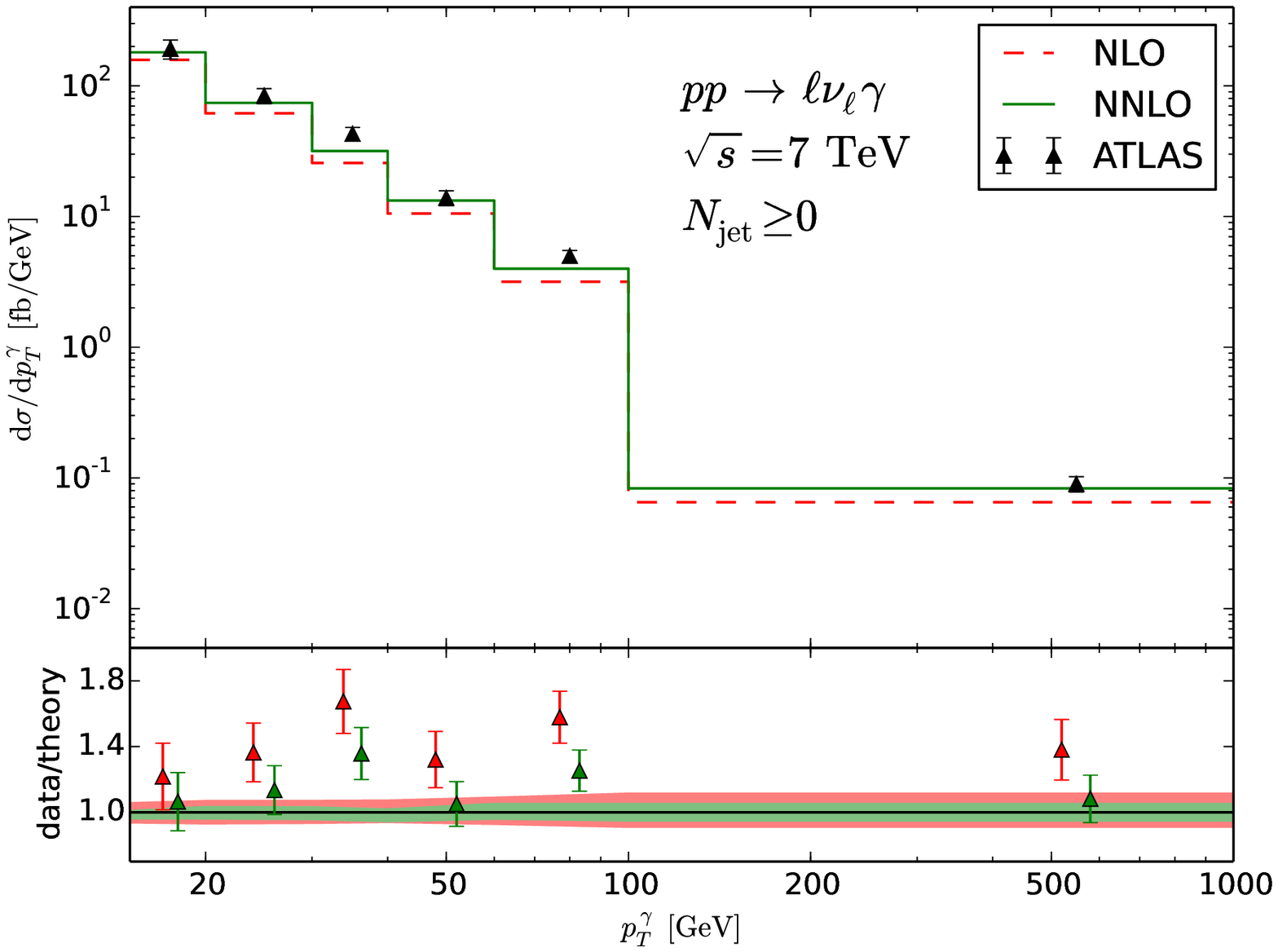}}
\\[-13em]
{\includegraphics[bb=18 190 594 590,clip,width=0.5\textwidth]{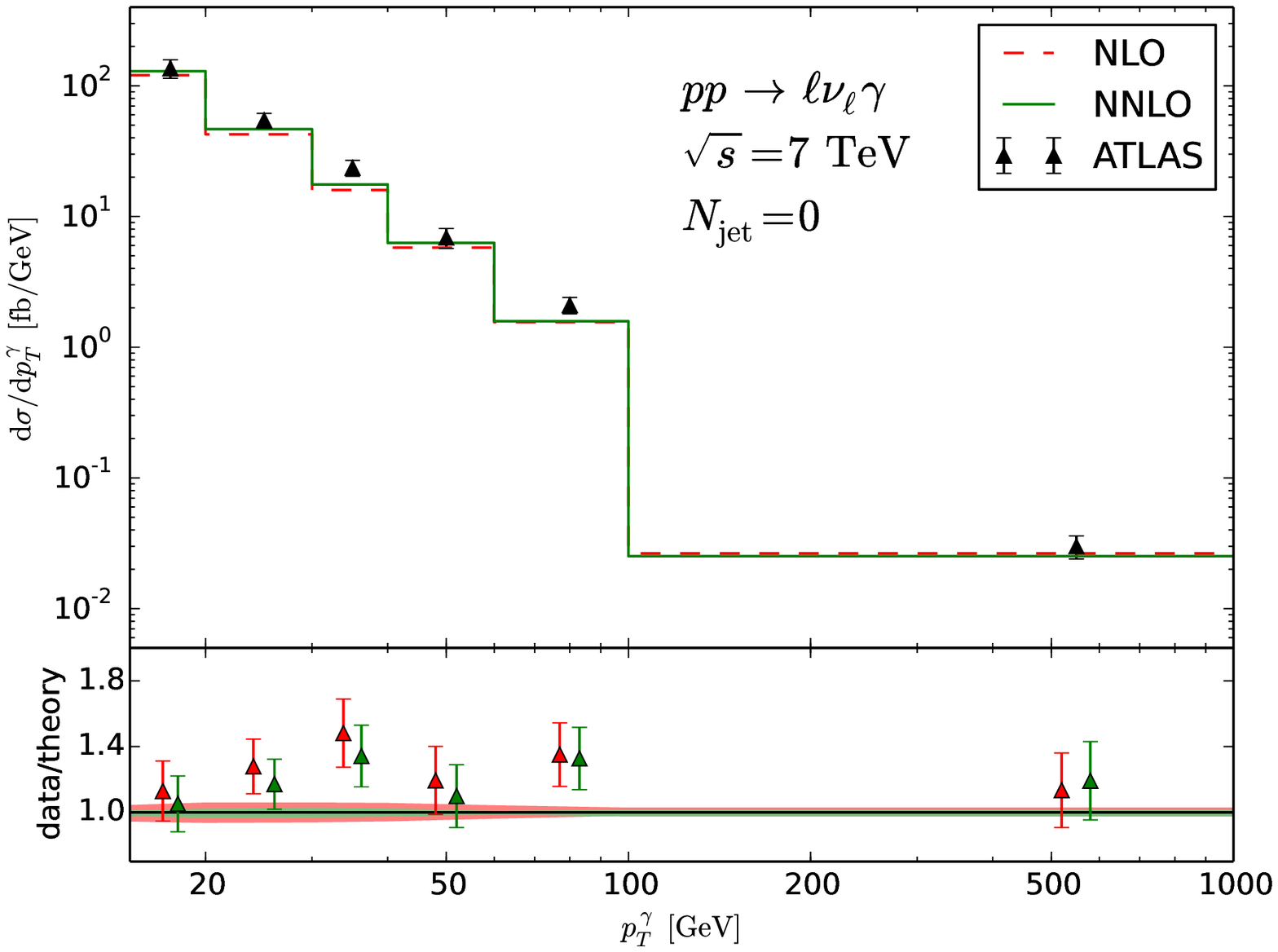}} 
\hspace{0em}
{\includegraphics[bb=90 250 455 734,clip,width=0.5\textwidth]{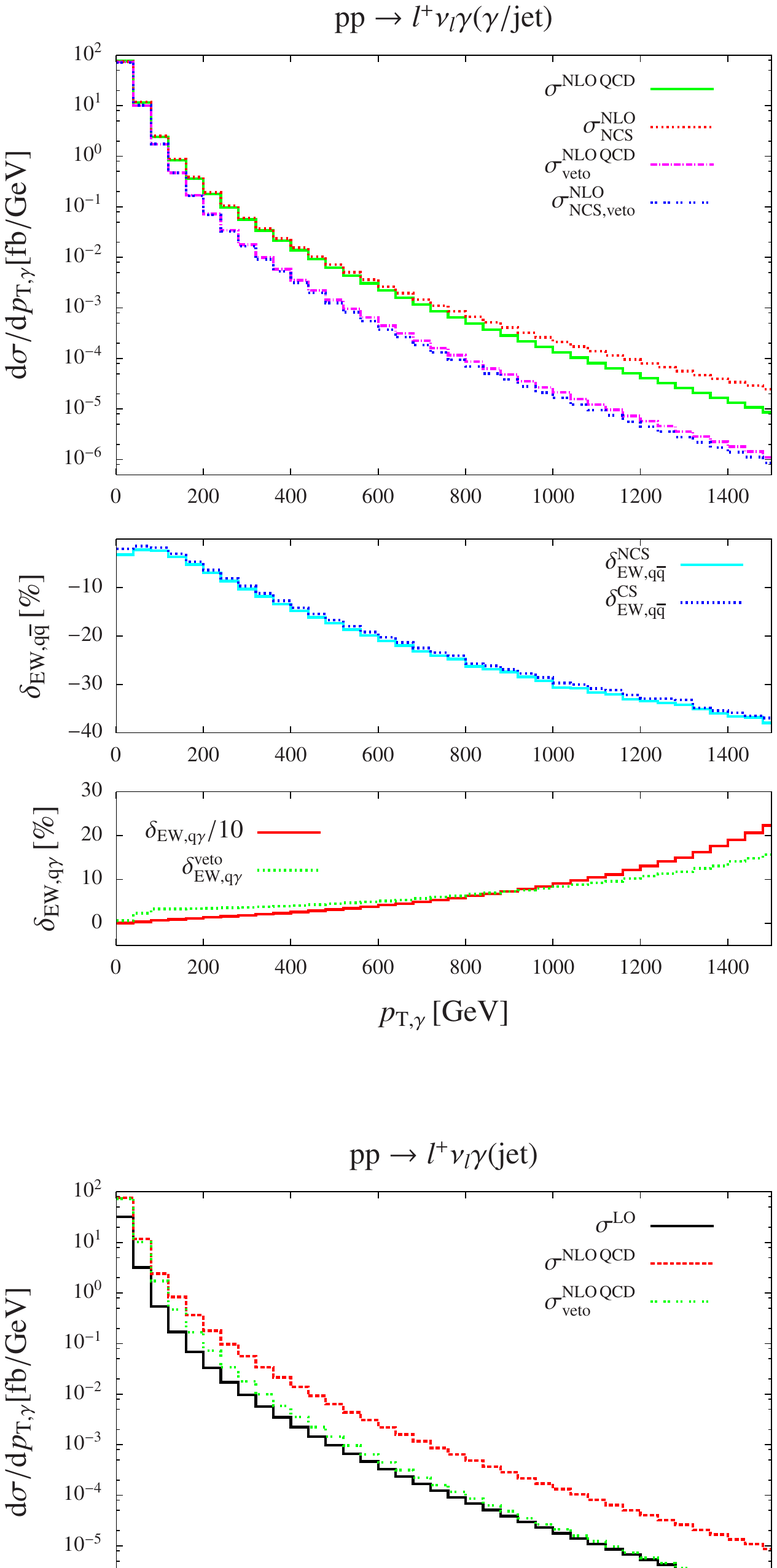}} 
\caption[]{Transverse-momentum distribution of the photon in
$\PW^++\gamma$ production at the LHC at NLO and NNLO confronted
with LHC data (left) and including various types of NLO
EW corrections (right)
(taken from Grazzini et al.~\cite{Grazzini:2015nwa} and Denner et al.~\cite{Denner:2014bna},
respectively).
}
\label{fig:Wgamma}
\end{figure}

The l.h.s.\ of Fig.~\ref{fig:Wgamma} shows a confrontation of
such a prediction for $\PW\gamma$ production with ATLAS data collected
at an energy of $7\TeV$. 
Already these data, which reach only up to few $100\GeV$ in the
hard-photon transverse momentum, favour the NNLO over the NLO prediction,
but upcoming data from LHC Run~2 will deepen the reach in
$p_{\mathrm{T}}^\gamma$ enormously with lower statistical errors,
leading to a more stringent test of the predictions.
At that level, the NLO EW corrections will become significant as well.
They are known for 
$\PZ\gamma/\PW\PW/\PW\PZ/\PZ\PZ$~\cite{Hollik:2004tm}
with stable W/Z bosons,
for $\PZ/\PW\gamma$~\cite{Accomando:2005ra}
and
$\PW\PW$~\cite{Billoni:2013aba} 
in leading pole approximation~%
\footnote{Predictions for the production of massive decaying gauge-boson pairs
based on approximated NLO EW corrections are included in 
{\sc Herwig}~\cite{Gieseke:2014gka}.},
and for 
$\PW\gamma$~\cite{Denner:2014bna}
fully including decay and off-shell effects.
The r.h.s.\ of Fig.~\ref{fig:Wgamma} shows results from the latter calculation of the NLO
EW corrections, demonstrating again large negative EW corrections 
for increasing $p_{\mathrm{T}}^\gamma$.
The bottom panel additionally reveals a surprisingly large impact
of photon-induced collisions from the partonic process
$q\gamma\to\PW\gamma q$ (with $q$ any quark or antiquark), even
in the presence of a jet veto.
Since the photon density is rather uncertain ($\sim100\%$) for large parton
momentum fractions~\cite{Ball:2013hta},
an improved determination of the photon
density is necessary to further stabilize $\PW\gamma$ predictions in the TeV range.

\section{Higgs-boson production}

Predictions for the various production channels of Higgs bosons were
continuously improved and refined in the last two decades.
The basic concepts and results were collected and reviewed in the
last years by the LHC Higgs Cross Section Working 
Group~\cite{Dittmaier:2011ti}
and in review articles such as Ref.~\cite{Dittmaier:2012nh}.
Here we restrict ourselves to a short discussion of the production
of a single Higgs boson with or without a hard jet.

\subsection{Single-Higgs-boson production}

\begin{figure}
\includegraphics[bb= 42 80 410 750,clip,width=0.25\textwidth,angle=90]{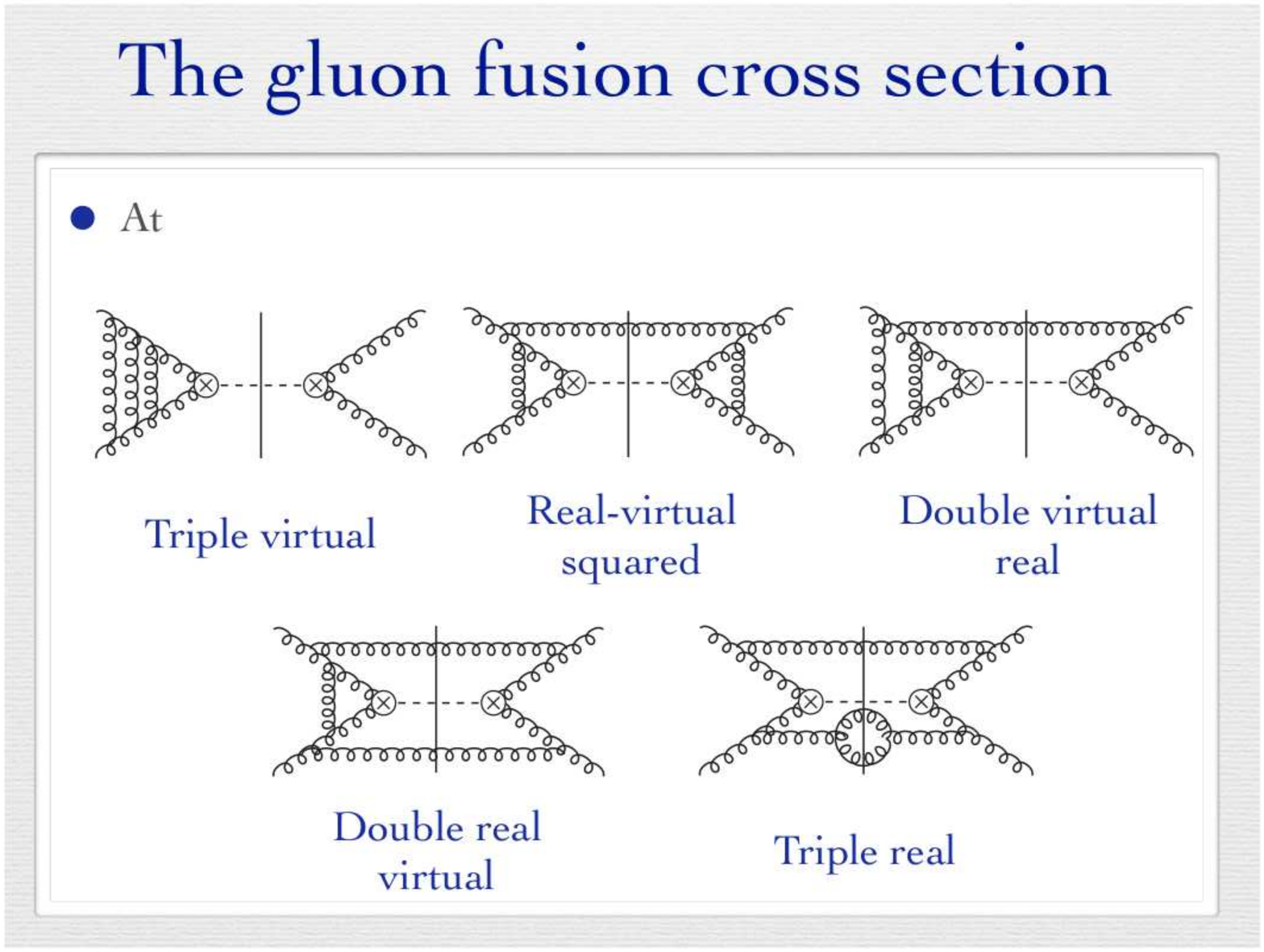} 
\hspace{1em}
\includegraphics[width=0.5\textwidth]{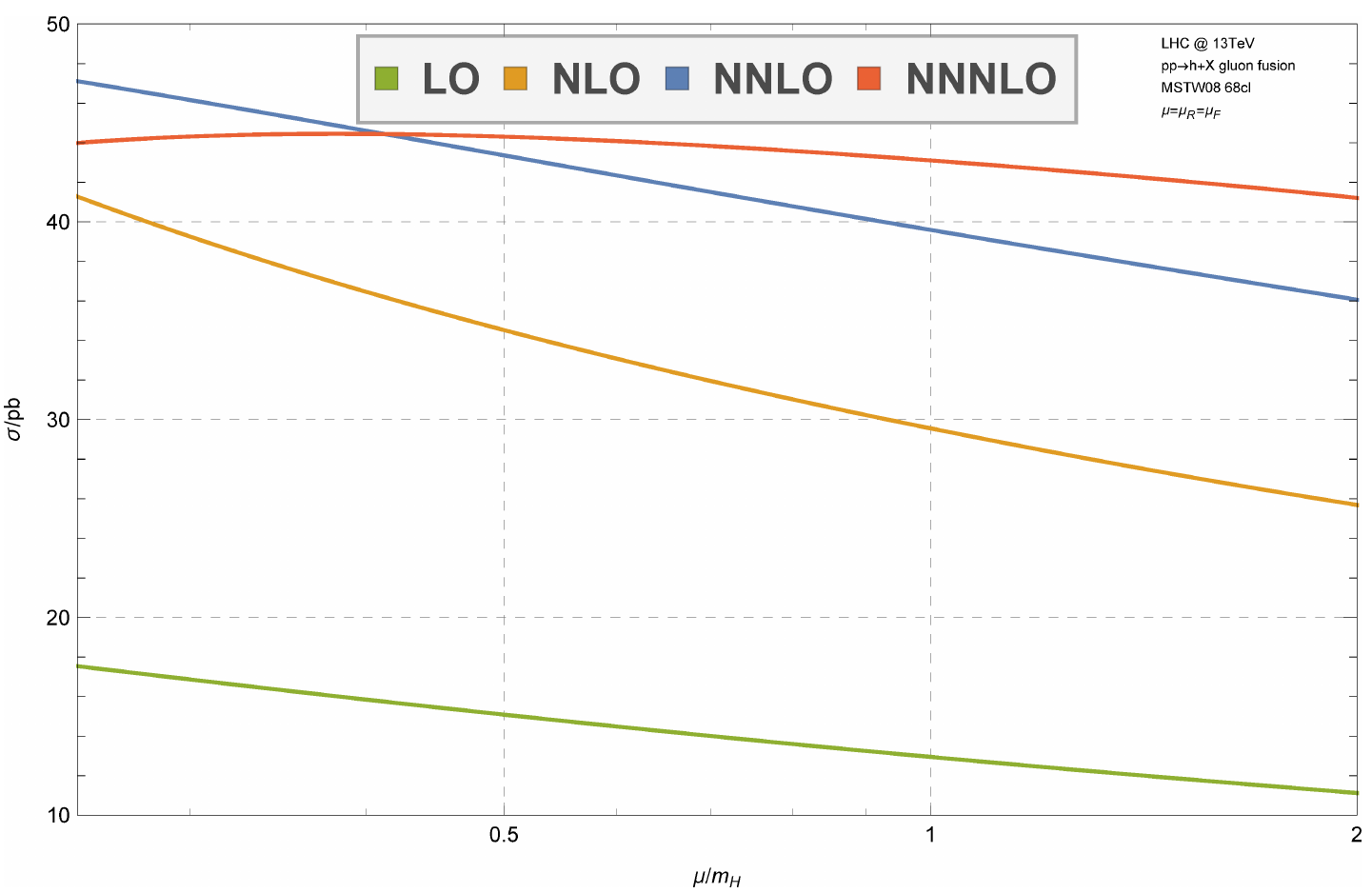} 
\caption[]{Relevant diagrammatic structures (left) for NNNLO QCD corrections
to Higgs production via gluon fusion, $\Pg\Pg\to\PH$,
and scale dependence ($\mu=\mu_{\mathrm{ren}}=\mu_{\mathrm{fact}}$)
of the corresponding $\Pp\Pp$ cross section
at LO, NLO, NNLO, and NNNLO
(taken from Anastasiou et al.~\cite{Anastasiou:2015ema}).
}
\label{fig:Hnnnlo}
\end{figure}

In the SM
single-Higgs-boson production at hadron colliders is a loop-induced process
where mainly two gluons produce the Higgs boson via a top-quark loop.
The mass hierarchy $2\Mt\gg\MH$ and the fact that the relevant partonic
scattering energy $\sqrt{\hat s}$ concentrates in the vicinity of the threshold
at $\MH$ implies the possibility to evaluate transition amplitudes in terms
of an asymptotic series in inverse powers $1/\Mt^n$ of the top quark.
This approach can nicely be formalized in the framework of an effective field
theory with local Higgs--gluon interactions where the top quark is integrated
out. While the $\Mt$-dependence of amplitudes and cross sections are known
to NLO for a long time~\cite{Graudenz:1992pv}
and to NNLO~\cite{Harlander:2009mq} QCD
up to the relevant powers in $1/\Mt$, the
leading contribution (which is of $1/\Mt^0$, i.e.\ a constant in $\Mt$)
to the cross section has been recently worked out even to the NNNLO QCD 
level~\cite{Anastasiou:2014lda,Anastasiou:2015ema} (see also Ref.~\cite{mistlberger}),
which documents the first cross-section calculation at this order.
The l.h.s.\ of Fig.~\ref{fig:Hnnnlo} shows the relevant types of
interference terms contributing to the cross section at this level,
whose calculation is of enormous complexity.
The integrated partonic cross section is calculated as asymptotic series
organized in powers $(1-z)^n$, where $z=\MH^2/\hat s$, i.e.\
virtual and soft-gluon contributions appear at threshold where $z\to1$.
In order to achieve sufficient precision,
about 35~terms where required in this series expansion.
The scale dependence of the resulting $\Pp\Pp$~cross section goes down from
$\sim9\%$ to $\sim3\%$ in the step from NNLO QCD to NNNLO QCD, as
illustrated on the r.h.s.\ of Fig.~\ref{fig:Hnnnlo}, which nicely
shows the perturbative convergence from LO through NNNLO.
The NNNLO correction itself is $\sim2\%$ at the scale $\mu=\MH/2$.
It is not yet fully understood why the approximate NNNLO 
result~\cite{Bonvini:2014jma} based on virtual+soft corrections at
$z\to1$ in combination with the known high-energy behaviour at
$z\to0$ does not agree with the full calculation very well.

Another issue in this context concerns the proper estimate of the
full theoretical uncertainty which is, e.g., delicate because
parton distribution functions extracted from experiment at NNNLO
are not available.
Moreover, several other effects play a role
at this level of precision of few percent, 
such as the influence of a finite bottom mass (only known to NLO)
and the combination with the NLO EW corrections~\cite{Actis:2008ug},
which amount to $5\%$ and whose factorization is based on 
arguments motivated by an effective field theory for 
$\MH\to0$~\cite{Anastasiou:2008tj}.

\subsection{Higgs-boson production in association with a hard jet}

\begin{figure}
\includegraphics[width=0.5\linewidth]{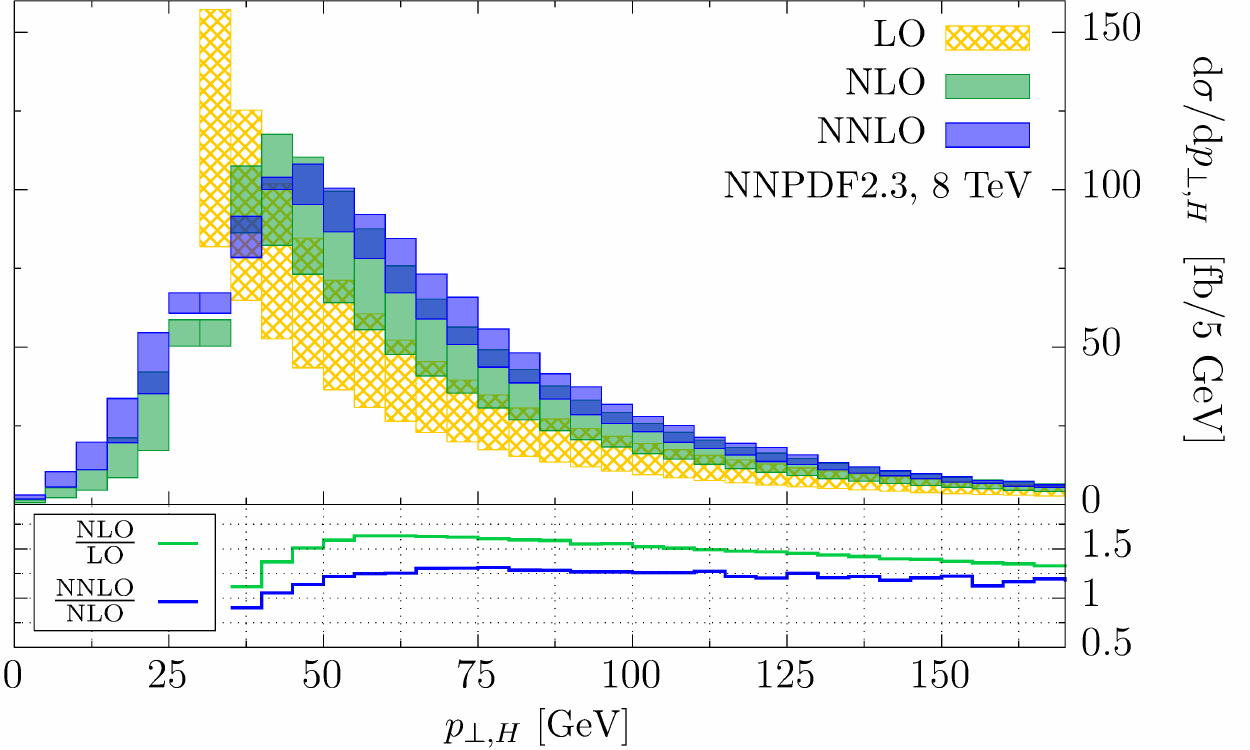}
\hspace{1em}
\includegraphics[width=0.5\linewidth]{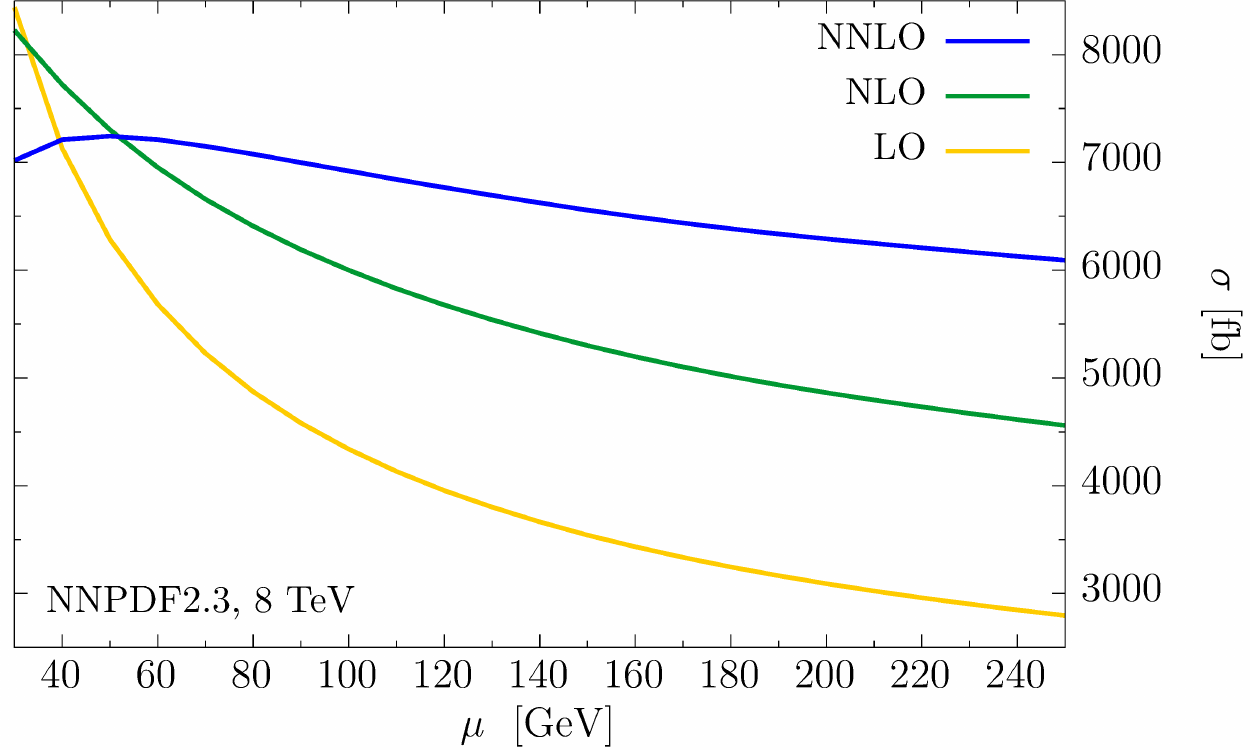}
\caption[]{Prediction for the transverse-momentum spectrum of Higgs-boson production
at the LHC at LO, NLO, and NNLO QCD
and corresponding $K$~factors (left),
and scale dependence (right) of the corresponding integrated cross section 
(taken from Boughezal et al.~\protect{\cite{Boughezal:2015dra}}).
}
\label{fig:Hjet}
\end{figure}

The transverse-momentum spectrum of the Higgs boson is an important
observable to identify and analyze Higgs-boson events.
Since single-Higgs-boson production proceeds via
$\Pg\Pg\to\PH$ in LO, the Higgs boson can receive a non-vanishing
transverse momentum only via the radiation of a jet
(or other particles in higher orders).
Recently, the QCD predictions of the $p_{\mathrm{T},\PH}$
spectrum were pushed to the NNLO level~\cite{Boughezal:2015dra,Boughezal:2015aha},
where the IR singularities were treated with so-called
sector-improved residue subtraction~\cite{Czakon:2010td}
in the former calculation (see also Ref.~\cite{caola})
and with
jettiness subtraction~\cite{Boughezal:2015eha,Gaunt:2015pea} in the latter.%
\footnote{NNLO QCD results based on the 
$\Pg\Pg$~channel only~\cite{Chen:2014gva}
were also evaluated with the help of antenna subtraction.}
Figure~\ref{fig:Hjet} summarizes the central results of these calculations.
The QCD scale uncertainty of the integrated cross section
with $p_{\mathrm{T,jet}}>30\GeV$ (shown on the r.h.s.)
goes down from $\sim23\%$ at NLO to $\sim9\%$ at NNLO.
This behaviour is also observed for the relevant part of
the $p_{\mathrm{T},\PH}$~spectrum, shown in the l.h.s.\
of Fig.~\ref{fig:Hjet}, which reveals NNLO QCD corrections
of $\sim20\%$ on top of the NLO prediction.
NLO EW corrections to \PH+jet production, which involve
$2\to2$ two-loop amplitudes with many scales, are not
yet known, but their size most probably does not exceed the
size of the remaining QCD scale uncertainty for not too large
transverse momenta.

\section{Conclusion}

The field of perturbative calculations for particle collisions has developed very rapidly
in recent years towards higher and higher precision in many different directions.
The process of automating NLO calculations for multi-particle production reached
a rather mature state in QCD and is successfully ongoing on the EW side.
Beyond NLO, more and more results at NNLO QCD have become available for $2\to2$
scattering processes, and $2\to1$ processes, such as inclusive Higgs-boson production,
are even treated in NNNLO QCD.
Selected highlights at this frontier have been briefly discussed in this 
short article. 

Apart from the discussed directions, there was also significant progress in the
field of analytic resummations and parton-shower approaches and their combination
with fixed-order calculations, an issue that could not be touched here. 
The same applies to new concepts and techniques of more formal field theory
or mathematics as well as all kind of applications to physics beyond the SM, etc.
A short article can hardly do justice to the great success of the whole
field.

\section*{Acknowledgments}

I would like to thank the organizers for perfectly setting up the
conference at such an inspiring and interesting place like
Ch\^ateau Royal de Blois.

\end{document}